\documentclass[twocolumn,showpacs,preprintnumbers,amsmath,amssymb,floatfix,superscriptaddress,PRC]{revtex4}
\usepackage{graphicx}
\usepackage{dcolumn}
\usepackage{bm}
\usepackage{graphicx}

\begin{document}
\preprint{APS/123-QED}

\title{Measurements of neutron-induced reactions in inverse kinematics}

\author{Ren{\'e}~Reifarth}
\affiliation{Goethe-Universit\"{a}t Frankfurt am Main, Max-von-Laue-Str.1, 60438 Frankfurt am Main, Germany}
\author{Yuri A. Litvinov}
\affiliation{GSI Helmholtzzentrum f\"ur Schwerionenforschung, 64291 Darmstadt, Germany}
\affiliation{Max-Planck-Institut f\"ur Kernphysik, 69117 Heidelberg, Germany}

\date{\today}

\begin{abstract}

Neutron capture cross sections of unstable isotopes are important for 
neutron induced nucleosynthesis as well as for technological applications.
A combination of a radioactive beam facility, an ion storage ring and a high flux reactor
would allow a direct measurement of neutron induced reactions over a wide energy range 
on isotopes with half lives down to minutes. 

\end{abstract}



\maketitle

\section{Introduction}
\label{}

Knowledge of neutron-induced reaction rates is indispensable for 
nuclear structure and nuclear astrophysics 
as well as for a broad range of applications, where the obvious example of the latter is 
the reactor physics.
In nuclear astrophysics the neutrons with energies between 1~keV and 1~MeV play the most essential role since this energy range corresponds to temperature regimes relevant
for nucleosynthesis processes in stellar objects.

In this context $(n,\gamma)$ cross sections 
for unstable isotopes are requested for the $s$-process \cite{KGB11}, related to stellar helium 
burning, as well as for the $r$- \cite{TMP07} and $p$-processes \cite{ArG03}, related to explosive nucleosynthesis 
in supernovae. In the $s$-process, these data are required for analysing 
branchings in the reaction path, which can be interpreted as diagnostic tools for 
the physical state of the stellar plasma \cite{RAH03}. Most of the nucleosynthesis reactions during 
the $r$- and $p$-processes proceed through nuclides outside the stability valley, thus involving rather 
short-lived nuclei. Here, the challenge for $(n,\gamma)$ data is linked to the freeze-out 
of the final abundance pattern, when the remaining free neutrons are captured as 
the temperature drops below the neutron separation energy \cite{SuE01}. 
Since  many of these nuclei are too short-lived to be accessed by direct measurements \cite{CoR07} it 
is, therefore, essential to obtain as much experimental information as possible 
off the stability line in order to assist theoretical extrapolations of nuclear 
properties towards the drip lines. 

Apart from the astrophysical motivation there is continuing interest on neutron 
cross sections for technological applications, i.e. with respect to the neutron 
balance in advanced reactors, which are aiming at high burn-up rates, as well as 
for concepts dealing with transmutation of radioactive waste \cite{CHO11,BBE13}.

In general, the shorter the half life of the isotopes under investigation, the 
more difficult it will be to prepare a radioactive sample, to place the sample close to a detector
and to perform a reaction measurement. For proton and $\alpha$-induced reactions, one
solution to this problem is to invert the kinematics, namely to employ a radioactive ion beam hitting a proton or helium 
target at rest. 
In the case of neutrons, this approach would require a neutron target, where the neutrons are themselves unstable. 
In this article, we describe a possible solution to this problem, which allows direct measurements of
neutron-induced cross sections on radioactive isotopes.
Nuclei with half-lives as short as a few minutes or even below can be studied. 
Such short-lived isotopes can currently not be directly investigated. 

\section{Concept}
\label{scheme}

\begin{figure}
  \includegraphics[width=.4\textwidth]{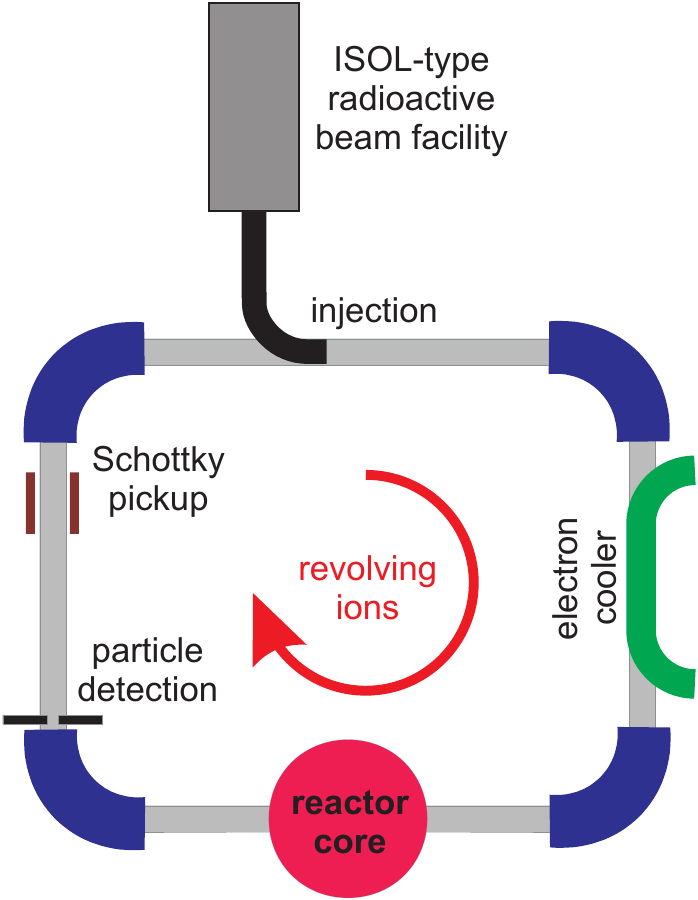}
  \caption{Schematic drawing of the proposed setup. 
  Shown are: the facility to produce and separate exotic nuclei of interest (dark gray) and
  the main components of an ion storage ring which include the beam lines and focusing elements (gray), 
  dipoles (dark blue), electron cooler (green), an intersected reactor core (red), 
  particle detection capability (black) and Schottky pick-up electrodes (brown).
  \label{setup}}
\end{figure}

The idea presented in this article is to measure neutron-induced reactions on radioactive ions in inverse kinematics. 
This means, the radioactive ions will pass through a neutron target. 
In order to use efficiently the rare nuclides as well as to enhance the luminosity,
it is proposed to store the exotic nuclides in an ion storage ring.
The neutron target can be the core of a research reactor, 
where one of the central fuel elements is replaced 
by the evacuated beam pipe of the storage ring. Such geometries are fairly common for research reactors, in particular of TRIGA type. At least one fuel element is typically replaced by an pipe, which can be loaded with capsules containing sample material to be irradiated. Another possibility would be to add the beam pipe right next to the reactor core. The core is usually surrounded be water and -in the case of research reactors- moderating water. The neutron densities at the edge of the reactor core are typically one order of magnitude smaller than in the center of the core. This can be taken into account when planning such a setup. An evacuated beam pipe as needed for the proposed setup would therefore not interfere with the operation of the reactor. More demanding might be the vacuum required for the operation of the storage ring. Usually, it is necessary to bake the corresponding structural parts in order to achieve UHV-conditions. However, it has been shown that proton-induced reaction can be investigated in this kinematics \cite{ZAB10}. The revolving ions were penetrating a hydrogen jet-target \cite{KPW09}, which has most likely a much bigger impact on the vacuum conditions than the walls.
The neutrons can easily penetrate the beam pipe creating a kind of neutron gas, which the revolving ions have to pass. 
The neutron density in the target is then only dependent on the power output of the reactor and the temperature of the reactor core.
A schematic drawing of the proposed setup is given in FIG.~\ref{setup}. 
The scheme is quite flexible, but for practical reasons, 
we base our discussion on the parameters of the existing rings. A dedicated design 
study could be performed in the future.

Two storage ring facilities are presently in operation which offer stored exotic nuclides.
These are the Experimental Storage Ring  (ESR)~\cite{Fra87} at GSI in Darmstadt 
and the experimental cooler-storage ring (CSRe)~\cite{XZW02} at IMP in Lanzhou.
Although these rings are capable of slowing stored ions down to a few AMeV, 
these rings are primarily designed to operate at energies around 200-400 AMeV~\cite{LGK10}.
Therefore we consider a storage ring similar to the Test Storage Ring (TSR)~\cite{GLR12} 
which was in operation until 2013 at the Max-Planck Institute for Nuclear Physics in Heidelberg.
Although this storage ring was not used to store radionuclides, there is a detailed technical design report to move TSR to CERN where
it shall be coupled to the ISOLDE radioactive ion beam facility~\cite{GLR12}.
The example parameters of the TSR are used in the following discussion. Even lower beam energies can be realized at cryogenically cooled rings, like the CSR \cite{HBB11} or the CRYRING \cite{HHS13}.

\begin{figure}
  \includegraphics[width=.49\textwidth]{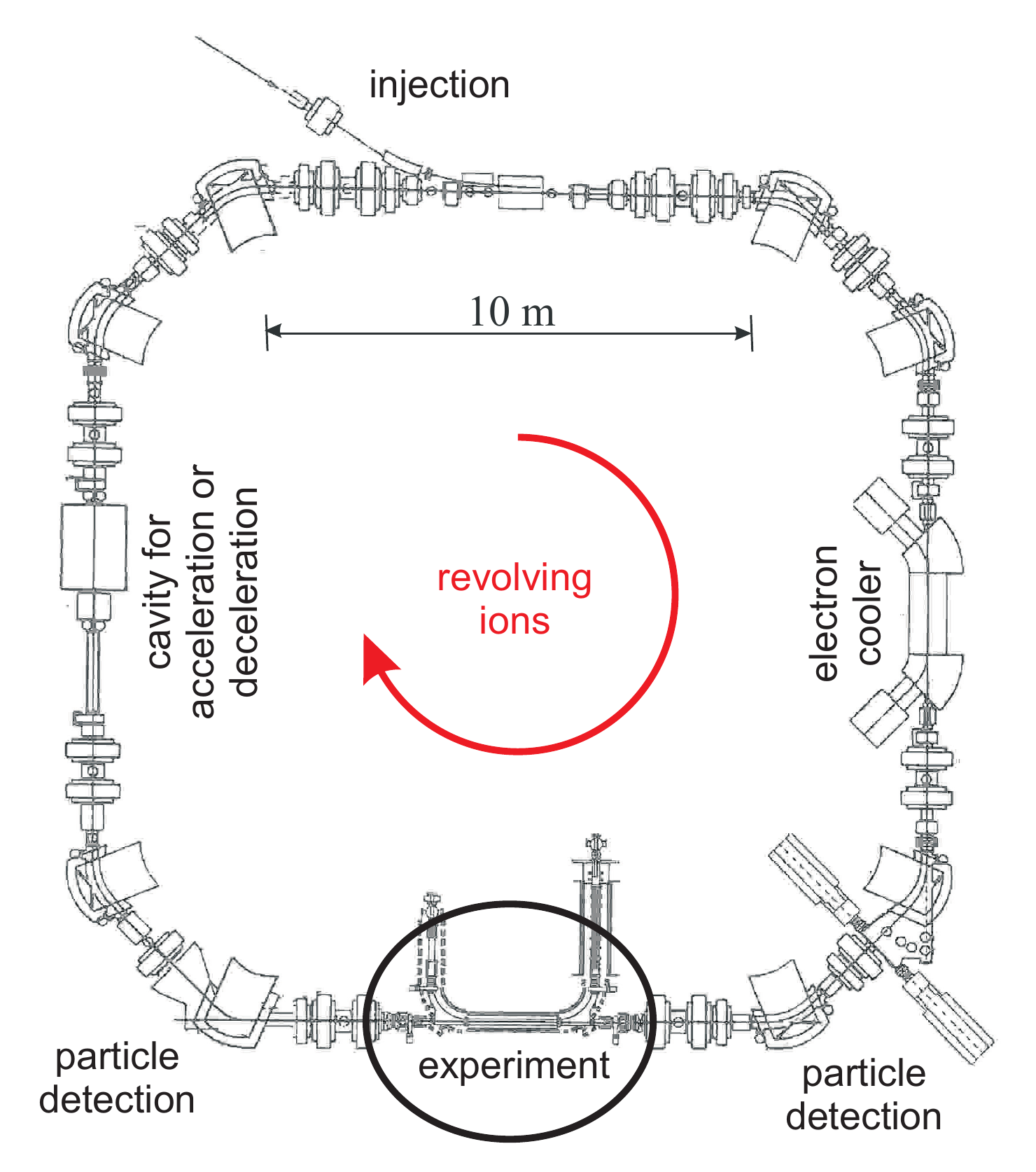}
  \caption{Schematic drawing of the Test Storage Ring, TSR~\cite{GLR12}. 
  The injection, electron cooler, acceleration cavity and particle detector setups are indicated.
  A core of a reactor can be imagined at place for in-ring experiments.
  Adopted from Ref.~\cite{GLR12}.
  \label{tsr}}
\end{figure}

The proposed storage ring shall have 4 straight sections as in the case of the TSR (see FIG.~\ref{tsr}).
One of these straight sections will go through the core of the reactor.
The ion-optics of the ring will be set such that the beam size is as small as possible in this section.
The size of the beam in horizontal $(x)$ and vertical ($y$) directions is given as $x=\sqrt{\beta_x\epsilon}$ and $y=\sqrt{\beta_y\epsilon}$, 
where the beta functions $\beta_x$ and $\beta_y$ describe the envelope of the beam in $x$ and $y$ directions, 
respectively, along the beam axis, and $\epsilon$ is the beam emittance.
Furthermore, in order to minimise the influence of the momentum distribution on the beam size, 
the dispersion function has to be minimal in this section.
Sufficiently small beta and dispersion functions are achieved, e.g., in the cooler section 
in the standard operation mode of the TSR  (see Figs.~34 and 36 in Ref.~\cite{GLR12}).
In this case the size of the beam will be the smallest (waist) in the middle of the reactor core.
The size of the TSR beam pipe is 20 cm in diameter which can be taken here as a maximum size.

One other straight section will be taken by the electron cooler. 
This is an essential component of the setup.
The electron cooling is needed to achieve and to keep a small beam emittance, 
which is defined by the equilibrium between the cooling force on the one side and 
effects acting against it, like intra-beam scattering or energy losses in the rest gas, on the other side.

The TSR is equipped with rf-cavities to accelerate/decelerate the stored beams which is not needed 
for our setup as it is much more efficient to inject and to store the ions directly at the required energy. 

The straight section opposite of the neutron target will be occupied by the injection hardware.
The so-called multi-turn injection is employed at the TSR~\cite{GLR12} and is suggested to be used here.
In the multi-multi injection the horizontal acceptance of the storage ring is filled with ions. 
With a help of septa magnets the ions are injected into the ring for several tens of revolutions ($\sim150~\mu$s in case of the TSR).
Afterwards, the ions are electron cooled which compresses the phase-space and empties a part of the ring acceptance.
This emptied space can again be used to inject new ions. 
The electron cooling requires a few hundreds of milliseconds until the next multi-multi injection can be performed.
This so-called ``electron-cooling stacking'' can be repeated continuously 
until an equilibrium is reached between the number of ions lost from the ring and the number of injected ions.
There is an upper limit due to space charge effects.
The TSR holds a record by storing the 18~mA current of $^{12}$C$^{6+}$ ions, 
though a conservative limitation for a stable operation is around 1~mA.
However, in our case this limit might become applicable only for ions with long lifetimes and large production rates.
In the following we will assume a moderate intensity of stored ions of $10^7$ particles in the ring at any given time.
 
Essential issue are the losses of the ions from the ring. 
Compared to other storage-ring based reaction measurements, where the internal target is the major source of ion losses,
using neutrons as target means that no atomic reactions in the target have to be taken into account.
Therefore, the two main loss mechanisms are atomic charge exchange 
reactions of the ions with the residual gas atoms and electron capture in the electron cooler.
The residual gas pressure of the TSR is $4-6\cdot10^{-11}$~mbar.
Numerous measurements of the beam lifetimes exist in the TSR for different ions, ionic charge states, and energies~\cite{GLR12}.

An ISOL-type radioactive ion beam facility can be a source of exotic nuclei. 
ISOL-beams combine high intensity and good quality. In particular all the isotopes discussed in the applications can be produced with sufficient rates \cite{SKL07}.
The extracted low charged ions from the target will be trapped and charge-bred in an electron-ion beam trap/source, the
scheme realised presently at ISOLDE/CERN~\cite{GLR12}. 
The highly charged ions can be extracted and post-accelerated to the required energy by a linear accelerator and then injected into the ring.

Dependent on the momentum-over-charge change in the neutron-induced reaction, the daughter nuclei can stay within the storage ring acceptance or not. 
In the former case the number of daughter ions can be monitored by a non-destructive Schottky spectroscopy~\cite{GBB10, NHL11} or by using a sensitive SQUID-based CCC-detectors~\cite{VGN13}.
In the latter case the reaction products will be intercepted by particle detectors located behind the first dipole magnet downstream the reactor core (see Fig.~\ref{setup}).
The feasibility of this has been demonstrated in the ESR 
this has been demonstrated in the ESR by detecting the $^{97}$Ru recoil ions produced in the $^{96}$Ru$(p,\gamma)^{97}$Ru reaction \cite{ZAB10}.

The discussed concept requires the presence on site of a reactor and an ISOL radioactive beam facility. 
One of such locations could be the Petersburg Nuclear Physics Institute in Russia (PNPI), where a new-generation reactor, PIK \cite{Ser89}, 
is being constructed and an ISOL facility, IRIS, is in operation a few hundreds of meters away.

\section{Rate estimates}

The neutron flux through an arbitrary area in reactors ranges from 
$\phi_{neutron}=10^{13}$~/cm$^2$/s in small research reactors (TIRGA Mainz type~\cite{KKB08})
to about $\phi_{neutron}=10^{15}$~/cm$^2$/s or more in modern research reactors (FRM-II Munich \cite{ZSK06} or ILL Grenoble \cite{BCC06}). 
For the following rate estimates, we will simply assume an averaged neutron flux of 
$\phi_{neutron}=10^{14}$~/cm$^2$/s.
This results 
in a neutron density in any given volume of

\begin{equation}\label{eq_neutron_density}
  \rho_{neutron} =\frac{\phi_{neutron}}{v_{neutron}}
\end{equation}

where $v_{neutron}=2200$~m/s denotes the average velocity of neutrons in a thermal reactor. Assuming an interaction
zone with a length of $l=0.5$~m, the areal density of neutrons as seen by the passing ions is:

\begin{equation}\label{eq_neutron_areal_density}
  \eta_{neutron} =\rho_{neutron}\cdot l = \frac{\phi_{neutron}\cdot l}{v_{neutron}}
\end{equation}

which gives $\eta_{neutron}\approx 2\cdot 10^{10}$~cm$^{-2}$.

The number of ions passing the volume in a given time is about $I_{particle}=10^{13}$~s$^{-1}$
($10^7$ stored ions circulating in a ring with a revolution frequency of $\sim1$~MHz and ) resulting in a beam luminosity of 

\begin{equation}\label{eq_luminosity}
  L =\eta_{neutron}\cdot I \approx 2\cdot 10^{23} \frac{1}{\mbox{s cm}^2}
\end{equation}

and gives number a reaction rate of

\begin{equation}\label{eq_rate}
  R = L t \sigma
\end{equation}

or the number of counts per day:
\begin{equation}\label{eq_daily_rate}
  C_{daily} = 20\cdot \sigma\mbox{[mb]}
\end{equation}

which means, 
cross sections down to a few mbarns can be measured with sufficient 
statistics within one day. The beauty of the method that it is applicable to comparably short-lived
radioactive isotopes.
The half-life limit is mostly determined by the production rate at the radioactive
ion facility and the beam losses due to interactions with the rest gas in the ring.

\section{Possible reactions to be measured}
In order to discuss the possible reactions, which could be investigated with a setup as proposed here, it is important to understand the kinematics. The radius $(r)$ of a trajectory of a charged $(q)$ massive $(m)$ particle with velocity $(v)$ in a homogeneous, perpendicular magnetic field $(B)$ follows immediately form the Lorentz force:

\begin{equation}\label{eq_lorentz_force}
  q v B = \frac{m v^2}{r}
\end{equation}

hence

\begin{equation}\label{eq_radius_magnetic_field}
  r = \frac{m v}{q B} = \frac{p}{q B}
\end{equation}
Equation \ref{eq_radius_magnetic_field} is even valid  for relativistically moving particles, if $p$ and $m$ are relativistic variables. 
Compared to the revolving beam energy (energies above 0.1~AMeV), the neutrons (energies of 25~meV) can always be considered to be at rest. 
For the purpose of this paper, all channels can be viewed as a compound reaction. In a first step, a nucleus $X+n$ is formed and in a second step, particles or photons are emitted. This means, the momentum and the charge of the revolving unreacted beam $X$ and the compound nucleus $X+n$ are the same, which means, both species will be on the same trajectory. However, the velocity, hence the revolution frequency, is reduced by the factor $A/(A+1)$. If the revolving ions have charge $Z=q/e$ and mass $A=12\cdot m/m_{^{12}C}$ one finds then for the ratio of radii:

\begin{equation}
\frac{r_{D}}{r_{P}} = \frac{Z_P}{Z_{D}}\frac{p_{D}}{p_{P}}=\frac{Z_P}{Z_{D}}\frac{A_P}{A_P+1} \frac{A_{D}}{A_P},
\end{equation}
where indices $D$ and $P$ denote the produced daughter and unreacted parent nuclei, respectively.
And finally we obtain:
\begin{equation}\label{eq_ratio_radius_magnetic_field}
\frac{r_{D}}{r_{P}}=\frac{Z_P}{Z_{D}}\frac{A_{D}}{A_P+1}
\end{equation}

It depends now on the actual exit channel under investigation, which type of detection mechanism can and has to be applied.

\subsection{Neutron captures (n,$\gamma$)}
The neutron capture reaction can be viewed as a two-step process, where the neutron gets captured into a compound nucleus, which de-excites via $\gamma$-emission to the ground state.

\begin{equation}\label{eq_n_g}
  ^A_Z X +n \rightarrow ^{A+1}_{Z}X^* \rightarrow ^{A+1}_{Z}X + \gamma
\end{equation}

The compound nucleus has the same total momentum as the primary beam. 
However, the velocity, hence the revolution frequency, is reduced by the factor $A/(A+1)$. 
The product will receive a small relative momentum spread of less than $10^{-3}$ because of the $\gamma$-emission.  
The appearance  of the freshly synthesised ions can therefore be detected via this frequency change 
by applying the non-destructive Schottky detectors. It has been shown even single ions can be detected, even if the primary beam is still present in the ring \cite{SCL13}. 
A rather sharp line should appear in the Schottky spectrum, centred around $A/(A+1)$ 
with respect to the primary beam (the velocity, hence the frequency of the compound nucleus). 
The Schottky method was already successfully demonstrated at the ESR at GSI \cite{BFF96, LiB11}. 
Since neither the charge nor the momentum of the products are different from the unreacted beam, 
the neutron capture can not be detected with particle detectors, see also Equation~\ref{eq_ratio_radius_magnetic_field}.

\subsection{Neutron removals (n,2n)}
Similarly to the neutron capture reactions, also neutron removals at low energies can be viewed as a two-step
process:

\begin{equation}\label{eq_n_2n}
  ^A_Z X +n \rightarrow ^{A+1}_{Z}X^* \rightarrow ^{A-1}_{Z}X + 2n + \gamma
\end{equation}

The product ion has on average the same velocity as in the case of neutron capture, but the mass is reduced by two units. The momentum and hence the radius in a magnetic field of the produced ion compared to the 
unreacted beam is therefore reduced by (Equation~\ref{eq_ratio_radius_magnetic_field}):

\begin{equation}\label{eq_momentum_n_2n}
  \frac{r_{(n,2n)}}{r_{P}} = \frac{A_P-1}{A_P+1}
\end{equation}

In addition, the momentum spread is now in the order of $1/A$ and it remains 
a  matter of momentum acceptance of the storage ring 
($\approx\pm1.2\%$ at the ESR and $\approx\pm3\%$ at the TSR) and the isotope under investigation, 
if the measurement is feasible or not. 
Even for heavy nuclei, a rather broad line should appear in the Schottky spectrum, centred around $A/(A+1)$ 
with respect to the primary beam (the velocity, hence the frequency of the compound nucleus). 

\subsection{Neutron-induced charged particle reactions $(n,Z)$}
Neutron-induced reaction with charged particles in the exit channel (like $(n,p)$ or $(n,\alpha)$) are difficult to measure in conventional kinematics, even for long-lived or stable isotopes. 
The reasons are of technical nature: In order to detect the charged reaction products, they have to be able to leave the sample. 
This necessitates very thin samples, which, in combination with small cross sections, results in a very limited number of successful measurements with fast neutrons.

In inverse kinematics however, this difficulty does not exist, provided that it is possible to detect the produced ions. 
The ratio of radii in case of a $(n,p)$ reaction is (Equation~\ref{eq_ratio_radius_magnetic_field}):

\begin{equation}\label{eq_momentum_n_p}
  \frac{r_{(n,p)}}{r_{P}} = \frac{Z_P-1}{Z_P}\cdot \frac{A_P}{A_P+1}
\end{equation}

And for the (n,$\alpha$) reaction:

\begin{equation}\label{eq_momentum_n_z}
  \frac{r_{(n,\alpha)}}{r_{P}} = \frac{Z_P-2}{Z_P} \frac{A_P-4}{A_P+1}
\end{equation}

This means, the separation of the trajectories of the products from the unreacted beam is huge and therefore the 
products can easily be detected with charged-particle detectors 
placed at a place with a large dispersion outside the trajectory of the unreacted beam.

The emission of the massive (charged) particle occurs isotropically in the center of mass system. Therefore the recoil of the product leads to a momentum spread - longitudinal as well as transversal. The momentum of the product $p_{P}^{cm}$ in the center of mass is:

\begin{equation}
  p_{P}^{cm} = \sqrt{2 \mu E_{kin}^{exit} }
\end{equation}

where $\mu=mM/(m+M)$ denotes the reduced mass. $E_{kin}^{exit}$ denotes the kinetic energy released in the center of mass and can have any value between zero and the some of the mass difference and the initial kinetic energy in the center of mass:

\begin{equation}
  0 \le E_{kin}^{exit} \le Q+E_{kin}^{initial}
\end{equation}

hence 

\begin{equation}
  p_{P}^{cm} \le \sqrt{2 \mu \left(Q+E_{kin}^{initial}\right) }
\end{equation}

The maximum relative momentum spread in the laboratory system is therefore:

\begin{equation}
  \frac{\Delta p_P^{cm}}{p_P} =  \sqrt{\frac{ \mu }{M} \frac{1}{A}\frac{Q+E_{kin}^{initial}}{E_{kin}^{initial}} } 
\end{equation}

If $m<<M$, the equation can be simplified to:

\begin{equation}
  \frac{\Delta p_P^{cm}}{p_P} =  \sqrt{\frac{ a }{A^2} \frac{Q+E_{kin}^{initial}}{E_{kin}^{initial}} } 
\end{equation}

If one considers a typical case for a $(n,p)$ reaction ($a=1$), a sample mass of $A=100$, $E_{kin}^{initial}=100$~keV (corresponding to 0.1~AMeV beam energy) and a mass difference of $Q=1$~MeV, the relative momentum spread is 3\%. This means, after 1~m of flight path, all the products would still be within a 3~cm radius around the primary, unreacted beam. The detection could be done for instance with a pair of particle detectors even before the next magnet. The detectors should be arranged such that they form a slit in the plane of the storage ring with the unreacted beam in the center of the slit. This allows the optimization of the primary beam during injection without interference with the detectors. 

If the Q-value is negligible, the relative momentum spread in the case of $(n,p)$ is 1/$A$, while it is 2/$A$ in the case of $(n,\alpha)$. In this case, the products should be separated from the primary beam in the field of the following dipol magnet.

\subsection{Neutron-induced fission $(n,f)$}

As with the previously discussed reactions, also neutron-induced fission can be viewed as a process with several steps. At first, a compound nucleus is created. Secondly fission occurs, where the nucleus typically splits into two smaller nuclei. In a third process, prompt neutrons are evaporated off the fission products. At last, comparably slowly, the fission products $\beta^-$-decay towards the valley of stability. The kinetic energy of the isotropically emitted fission products in the center of mass is $\approx 1$~AMeV. This means, at low beam energies, the fission products are emitted in all directions in the laboratory system and therefore difficult to detect with well-determined efficiency. At higher beam energies, the fission products become again focused in beam direction, and have a smaller $A/q$ ratio
than the primary beam. It would therefore be possible to detect the reaction products with charged particle detectors positioned just outside the primary beam. The determination of fission cross sections in inverse kinematics has been successfully proven, but not for neutron-induced fission \cite{BPA10}. The chain of $\beta^-$-decays will appear after the detection of the particles inside the detectors can be used for further investigation or discrimination against background.

In order to estimate the momentum spread, we assume the same mass for each the fission product. With a mass of $A=250$, a mass difference of $Q=190$~MeV and a beam energy of 10~AMeV (corresponding to a neutron energy of 10~MeV), the maximal momentum spread would be:

\begin{equation}
  \frac{\Delta p_P^{cm}}{p_P} =  \sqrt{\frac{ 1 }{2 A} \frac{Q+E_{kin}^{initial}}{E_{kin}^{initial}} } = 20\%
\end{equation}

If the beam energy is sufficiently high and the Q-value can be neglected, the maximal relative momentum spread is 1/$\sqrt{2A} \approx 5\%$. Depending on the geometry, a setup with 2 particle detectors with a slit in the plane of the ring can cover almost 100\% of the products.

\section{Applications}
\subsection{Nuclear astrophysics}
\subsubsection{The $s$-process}
About 50\% of the element abundances beyond iron are produced via slow neutron capture 
nucleosynthesis ($s$ process) \cite{KGB11}. Starting at iron-peak seed, the $s$-process mass flow 
follows the neutron rich side of the valley of stability. 
If different reaction rates 
are comparable, the $s$-process path branches and the branching ratio reflects the 
physical conditions in the interior of the star. Such nuclei are most interesting 
because they provide the tools to effectively constrain modern models of the stars 
where the nucleosynthesis occurs. As soon as the $\beta^-$ decay is faster than the 
typically competing neutron capture, no branching will take place. Therefore 
experimental neutron capture data for the $s$-process are only needed if the respective 
neutron capture time under stellar conditions is similar or smaller than the $\beta^-$ 
decay time, which includes all stable isotopes. Depending on the actual neutron density 
during the $s$-process, the "line of interest" is closer to or farther away from the 
valley of $\beta$-stability. 

\begin{figure}
  \includegraphics[width=.49\textwidth]{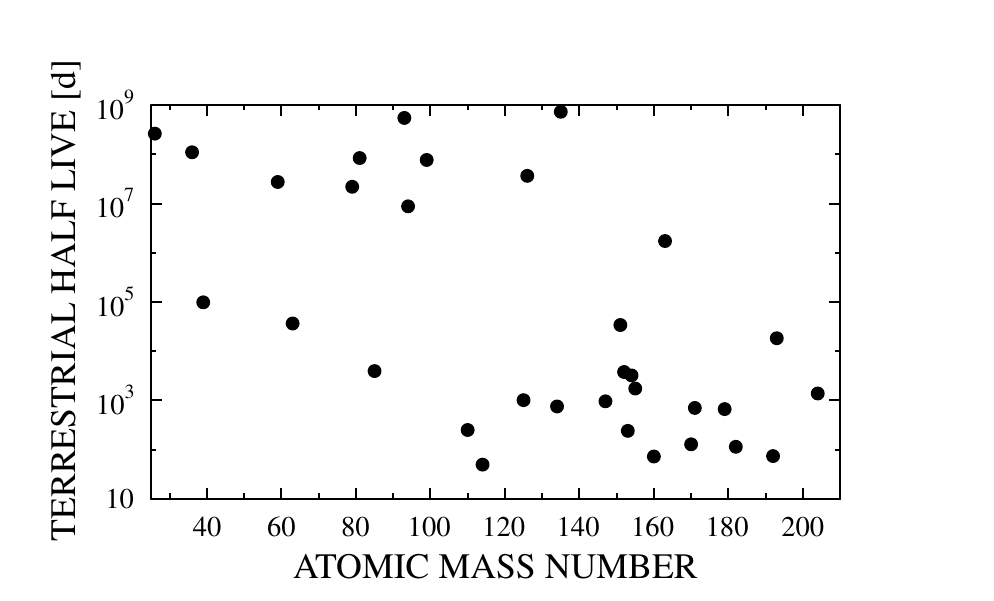}
  \caption{Terrestrial $\beta^-$ live times 
  for unstable isotopes on the classical $s$-process path as 
  a function of mass number. Shown are only isotopes where the neutron capture under stellar conditions
  is faster than the
  stellar $\beta^-$ decay for a neutron density of 4$\times 10^{8}$~cm$^{-3}$ at a temperature of 30~keV, the conditions of the classical $s$ process \cite{RAH03}. If the $\beta$-decay is faster than the neutron capture, the $s$-process proceed to the next higher element.
  \label{isotopes_ne22}}
\end{figure}

The modern picture of the main $s$-process component producing nuclei between iron and bismuth
refers to the He-shell burning phase in 
AGB stars \cite{LHL03}. The $s$ process in these stars experiences episodes of low neutron
densities of about $3\cdot10^7$~cm$^{-3}$, the $^{13}$C($\alpha$,n)-phase, and very high 
neutron densities, the $^{22}$Ne($\alpha$,n) phase.  
The highest neutron densities during the latter phase reach values of up to 10$^{11}$~cm$^{-3}$. 
FIG.~\ref{isotopes_ne22} shows a summary of the  
$\beta^-$ decay times for radioactive isotopes on the neutron rich side of the 
valley of stability, for the conditions during the main component of the classical $s$ process, which is in between the two phases of the $s$ process in AGB stars \cite{RAH03}. 
During the $^{22}$Ne($\alpha$,n) phase, the lifetime versus neutron capture is much shorter
resulting in isotopes with half-lives of just a few days forming the critical branching
points for the $s$-process reaction flow.

Because of the smaller total neutron exposure, the mass flow during the weak component of 
the $s$-process does not overcome the isotopes along the neutron shell closure of $N=50$
and is therefore restrictes to the mass region between iron
and yttrium. It takes place during convective core-He burning in massive stars
($M > 8 M_\odot$) and the material is later reprocessed during a second neutron exposure during convective
carbon shell burning of massive stars \cite{HKU07,RGB93,PGH10}. During the high temperatures of the carbon
shell burning of $T_9 \approx 1$ high neutron densities of up to $10^{11}-10^{12}$~cm$^{-3}$ can be reached,
similar to the conditions during the $^{22}$Ne($\alpha$,n) phase during the helium flash in AGB stars.

The most crucial neutron-induced reaction during the $s$-process is the neutron capture reactions. 
TABLE~\ref{tab_s-branchings} gives a small selection of interesting branch point nuclei, 
where direct determination of the neutron capture cross section is desirable. 
Even though all of these isotopes are very close to the valley of $\beta$-stability, 
none of them can be investigated with current or upcoming neutron time-of-flight (TOF) facilities. The currently strongest TOF-facility used for measurements in the astrophysical energy regime is DANCE at the Los Alamos National Laboratory \cite{RBA04}. At the sample position about $3\cdot 10^5$~n/s/cm$^2$ are available between 10 and 100~keV. Upcoming facilities like FRANZ \cite{RCH09} or the upgrade of nTOF \cite{GTB13} are aiming at neutron fluxes around $10^7$~n/s/cm$^2$. The last column of TABLE~\ref{tab_s-branchings} lists the minimum neutron flux 
in the keV regime necessary for a TOF measurement on the respective isotopes. It is obvious that even with the upcoming facilities, orders of magnitude are missing for a succesful measurement. However, the setup proposed here, would allow the corresponding measurement. The production of the corresponding nuclei in sufficient amounts is possible, since the isotopes are very close to the stability \cite{LGK06}.

\begin{table}[htb]
 \caption{Interesting branchpoint nuclei in the $s$-process nucleosynthesis network, which can not be directly 
 measured with current or upcoming facilities. 
 The last column gives an estimate for the minimum neutron flux necessary in the keV-regime 
 at the sample position for a traditional time-of-flight measurement using a 
 $4\pi$-calorimenter to detect the emitted $\gamma$-rays \cite{CoR07}.
         }
   \label{tab_s-branchings}
   \renewcommand{\arraystretch}{1.5} 
   \begin{ruledtabular}
   \begin{tabular}{ccc}
    Isotope      & half-life & neutron flux \\
                 & (d)      & (s$^{-1}$cm$^{-2}$) \\
    \hline
    $^{59}$Fe    & 45       & $10^{11}$\\
    $^{95}$Zr    & 64       & $10^{10}$\\
    $^{127}$Te$^m$    & 109       & $5\cdot 10^{9}$\\
    $^{147}$Nd    & 11       & $10^{9}$\\
    $^{148}$Pm$^{m, gs}$    & 41, 5       & $10^{9}$\\
		  		  
   \end{tabular}
   \end{ruledtabular}
\end{table}

If charged particles are in the exit channel of a neutron induced reaction, the experimental determination with traditional methods is restricted to a few favourable cases. Since the charged particle has to leave the sample, its thickness is very limited resulting in a correspondingly low reaction rate. The setup proposed here does not suffer from this requirement, since the resulting beam can easily be detected. Very important reactions are (n,$\alpha$) in particular on light nuclei, since they act as recycling points of the mass flow. Interesting isotopes are $^{33}$S, $^{36}$Cl, $^{37,39}$Ar, $^{40}$K and $^{41}$Ca \cite{RSK00}.

\subsubsection{The $r$-process}
The $r$-process synthesises roughly half the elements heavier than A=70. It proceeds through
neutron capture and beta decay \cite{CTT91a} at much higher neutron densities of $10^{20-22}$~cm$^{-3}$. 
Therefore capture is much faster than beta decay, so
that very neutron rich nuclei are created which decay back towards the valley of stability as the neutron density drops
marking the end of the $r$-process.
The neutrino-driven wind
model within core-collapse supernovae are currently
one of the most promising candidates for a successful $r$-process. 
These neutrino winds are thought to dissociate
all previously formed elements into protons, neutrons
and $\alpha$ particles before the seed nuclei for the $r$-process
are produced. Hence, the neutrino-driven wind model
could explain the observational fact that the abundances
of $r$-nuclei of old halo-stars are similar to our solar
$r$-process abundances \cite{TCP02}.

During the freeze-out phase, when the neutron density drops and the very short-lived nuclei decay, 
the $(n,\gamma)-(\gamma,n)$ equilibrium, which dominates 
the abundance distribution during most of the $r$-process episode, is interrupted. 
This means, neutron capture reactions can modify the final abundances. 
The sensitivity of the abundances in the $r$-process abundance 
peaks to changes in the neutron capture cross sections have been investigated \cite{SuE01}. 
The crucial reactions in range for the experimental setup proposed here are neutron captures 
on $^{130-132}$Sb with half lives between 2.8 and 40~min, 
and on $^{129-131}$Sn with half lives between 40~s and 7~min). 

\subsubsection{The $i$-process}

Under certain conditions, stars may experience convective-reactive nucleosynthesis episodes. 
If unprocessed, H-rich material is convectively mixed with an He-burning zone, it has been shown in 
hydrodynamic simulations that neutron densities in excess of $10^{15}$~cm$^{-3}$ can be reached \cite{HPW11, GZY13}. 
Under such conditions, which are between the $s$- and $r$-process, 
the reaction flow occurs a few mass units away from the valley of stability. 
These conditions are sometimes referred to as the $i$-process (intermediate process). 
One of the most important, but extremely difficult to determine rate, is the neutron capture on $^{135}$I. 
With an half-life time of around 6~h, this cross section can not be measured directly. With the setup proposed here, this cross section could be investigated directly, suffcient production yields of $^{135}$I provided.

\subsection{Advanced Reactor Technology}
The ratio of neutron capture to neutron induced fission is most important for the estimation of the criticality of nuclear reactors or other devices gaining energy via fission of heavy elements. 
Most experiments determining fission cross sections face the same problems like experiments determining reaction cross sections with charged particles in general. The short range of the charged fission products limits the sample thickness and therefore the reaction rate \cite{ToH07, THM09}. Experiments determining small neutron capture cross sections \cite{ERB08} in the presence of neutron induced fission have to deal with the $\gamma$-background following the deexcitation of the fission products \cite{JBB12}. All of those measurements are therefore usually only possible, if the investigated isotope is long-lived or stable. There are, however, a number of isotopes, which are very important, but short-lived.
Prominent examples of important isotopes are 
$^{235m}$U (25~min), $^{237}$U (6.75~d), $^{239}$U (23.5~min). These isotopes can currently not be investigated directly. The setups proposed here would allow measurements in an energy regime, which is not of immediate importance for currently operating reactors, since the average energy of the neutron spectrum is too low (thermal), but next generation reactors will not only operate at much higher temperatures, but the neutron spectrum will also be less moderated. This requires knowledge of the corresponding cross sections well into the MeV-regime. 


\section{Summary}
The combination of a modern storage ring with a neutron target consisting of a fission reactor allows the determination of a variety of neutron-induced cross sections in inverse kinematics. 
The luminosity of such a device would be sufficient to investigate cross sections down to millibarns 
on isotopes with half-lives down to a few minutes or may be even below. 
The energy in the centre of mass, which corresponds to the neutron energy in regular kinematics, 
depends on the isotope under investigation and can be as low as 100~keV. This energy limit is only given by the required life time of the beam, which might be improved in the future with new vacuum technologies.

\begin{acknowledgments}
We would like to thank F.~K\"appeler and K.~Sonnabend for many fruitful discussions. This work was 
partly supported by the HGF Young Investigators Project VH-NG-327, the Helmholtz-CAS Joint Research Group HCJRG-108, the GIF project G-1051-103.7/2009, the BMBF project 05P12RFFN6 and the EuroGenesis project MASCHE.

\end{acknowledgments}

\bibliography{/home/reifarth/Texte/paper/refbib}
\bibliographystyle{apsrev4-1}

\end{document}